# RELATIVISTIC GEOMETRY AND QUANTUM ELECTRODYNAMICS

**Gustavo González-Martín**

Departamento de Física, Universidad Simón Bolívar,

Apartado 89000, Caracas 1080-A, Venezuela.

Web page URL http:\\prof.usb.ve\ggonzalm\

Excitations of a relativistic geometry were used to represent the theory of quantum electrodynamics. The connection excitations and the frame excitations reduce, respectively, to the electromagnetic field operator and electron field operator. Because of the inherent geometric algebraic structure these operators obey the standard commutation rules of QED. If we work with excitations, we need to use statistical theory when considering the evolution of microscopic subsystems. The use of classical statistics, in particular techniques of irreversible thermodynamics, determine that the probability of absorption or emission of a geometric excitation is a function of the classical energy density. Emission and absorption of geometric excitations imply discrete changes of certain physical variables, but with a probability determined by its wave energy density. Hence, this geometric theory does not contradict the fundamental aspects of quantum theory.

# Introduction

A geometrical unified theory of gravitation and electromagnetism, which leads to equations of relativistic quantum mechanics, was shown to imply the quantization of electric charge and magnetic flux [1], providing a plausible explanation for the fractional quantum Hall effect. Furthermore the theory leads to a geometrical model for the process of field quantization[2], implying the existence of fermionic and bosonic operators. We shall indicate the latter paper by I. Here we discuss whether the theory gives quantum electrodynamics (QED) including its probabilistic interpretation.

A basic feature of the proposed theory is non-linearity. A solution cannot be obtained by the addition of two or more solutions and therefore it is not possible to build exact solutions from small subsystems. Nevertheless, it is possible to study its local linearized equations which represent excitations that evolve approximately under the influence of effects inherited from the nonlinear equation.

# Geometric Relations

### *Product of Jacobi Operators.*

The product $\alpha\beta$ has been taken in I as the Clifford product. Other candidates for the ring product were discussed in that paper. The chosen product presents difficulties when relating the geometry to standard QED.

Since the fiber bundle of frames, $E$, is a principal bundle and its vertical tangent bundle $TE$ has for fiber a Lie algebra structure inheritted from the group, it would be more natural that the chosen product be closed in the algebra so that the result be also valued in the Lie algebra. Geometrically we should specialize the ring product to be the Lie subproduct. Then the product $\alpha\beta$ is zero whenever its gradation is even. The bracket survives only when the commutator is nonzero because its gradation is odd and corresponds to the commutator. In other words the ring product obeys

$$\alpha \bullet \beta = [\alpha, \beta] \quad \text{for odd grade} , \qquad (1)$$

$$\alpha \bullet \beta = 0 \quad \text{for even grade} , \qquad (2)$$

with this product, the bracket defined in I becomes

$$\{\alpha V, \beta W\} = \{\beta W, \alpha V\} , \qquad (3)$$

which is the anticommutator, for the matter Jacobi operator fields.

The fiber bundle of connections, $W$, is an affine bundle and the ring product associated to the fiber of $TW$ is commutative. Therefore, the bracket is the commutator for the connection Jacobi operator fields

### *Commutation Relations*

If we take Schwingwer's action principle as described in I, we obtain the quantization relations by requiring that the quantum operators $\Psi$ be Jacobi operators,

$$\{\Psi(x), \delta\Psi(y)\} = 0 , \qquad (4)$$

$$\{\Psi(x), \Pi^\mu(y)\} = -\delta^\mu(x, y) . \qquad (5)$$

As discussed in previous articles, the Jacobi vectors $V_s$ represent fluctuations of sections of fiber bundles $E$. The corresponding jet prolongation $_jV$ of an extension of $V_s$ is a vector field on $JE$ which acts as an operator on functions on $JE$. A Jacobi field, as a vertical vector on the section $s(M)$ may be considered as a displacement of $s$. Similarly, its jet prolongation may be considered as a displacement of $^js$, in other words a variation,

$$\frac{d^js}{d\lambda} = {}_jV\left({}^js\right) . \qquad (6)$$

This linear action of the Jacobi fields on the background sections allow us to associate them with quantum theory objects. We may consider the background section as the state $|\rangle$ of the physical system and the Jacobi fields as physical linear operators $\Psi$ that naturally act on the states. In this condition the background section remains fixed and the Jacobi operators obey the first order linearized equations. (Heisenberg picture).

A given physical matter section may be expressed in terms of reference frame sections. This reference frames are also physical systems. Hence they may evolve under the action of the same group that acts on the physical sections. This represents the known equivalence of the active and passive views of evolution. We may consider the physical system evolving in a fixed frame or, equivalently, we may consider the system fixed in an evolving reference frame.

We may choose the reference frame so that the substratum section does not evolve. In this condition the substratum section remains fixed and the Jacobi operators obey the first order linearized equations of motion. (Heisenberg picture).

The Jacobi vector fields and their jet prolongations transform under the adjoint representation of the group that transforms the sections. We may use these transformations to a new reference frame where the Jacobi operators do not evolve. In this condition, the time dependence of the Jacobi operators may be eliminated and the substratum sections obey equations of motion, indicating that the state is time dependent (Schroedinger picture).

## Geometric Electrodynamics

### *Reduction of the electromagnetic sector*

In order to get the standard QED from the geometric theory, we have to reduce the structure group to one of the 3 U(1) subgroups in the SU(2) electromagnetic sector. The corresponding u(1) component of the fluctuation of the generalized connection is the electromagnetic potential *A*. Similarly, an electromagnetic fluctuation of the matter section determines a fluctuation of the generalized current. The standard QED electric current is the electromagnetic sector component of this current fluctuation,

$$j = \tfrac{1}{4}\operatorname{tr}(\kappa^1\kappa^2\kappa^3 \delta J) = \tfrac{1}{4}\operatorname{tr}(\kappa^1\kappa^2\kappa^3[J,\kappa^5]) = \tfrac{1}{4}\operatorname{tr}(\kappa^5\kappa^1\kappa^2\kappa^3 J) = \tfrac{1}{4}\operatorname{tr}(\bar{\kappa}^0 J) \quad . \quad (7)$$

It was shown in [3], that the standard electric current in quantum theory is related to the $\kappa^0$ component of the generalized current,

$$\tfrac{1}{4}\operatorname{tr}\bar{\kappa}^0 J^\mu = \tfrac{1}{4}\operatorname{tr}(\bar{\kappa}^0 \tilde{e}\kappa^\mu e) = \bar{\Psi}\gamma^\mu \Psi \quad , \qquad (8)$$

which is the electric current for a particle with a charge equal to one quantum in the geometric units.

It should be clear that an exact solution of the proposed problem is not possible with this technique, The reason for this is the presence of the nonlinearity of the self interaction in the equations. We must consider approximate solutions. In QED the interaction is imposed on special fields called "free fields". Here we have to discuss which frame and connection sections correspond to "free fields".

The field equation,

$$D^*\Omega = 4\pi\alpha J \quad , \qquad (9)$$

reduces when the connection has only a $\kappa^0$ electromagnetic component to

$$d^*dA = 4\pi\alpha j \quad . \qquad (10)$$

It is natural to consider a free connection field as one that satisfies the previous equation with *j=0*. A solution to the free equation provides a background connection section that can be used with the proposed technique.

The equation of motion

$$\kappa^\mu \nabla_\mu e = 0 \quad , \qquad (11)$$

presents the difficulty with the self interaction. It is not possible to set the connection to zero because it will automatically eliminate the self energy and the mass parameter in accordance with our theory. The simplest approximation is then to assume that all self interaction effects, to first order, are concentrated in the single mass parameter. The fluctuation equation is,

$$\kappa^\mu \partial_\mu e = me + \text{ interactions } \quad . \qquad (12)$$

It is natural to consider a free frame field as one that satisfies the previous equation without the interaction term. A solution to this free equation provides a background frame section that can be used with our technique. The linearized fluctuation equations, with the interaction term, determine the fluctuations of the connection and frame fields.

### *Quantum Electrodynamics*

We shall consider that the cause of the fluctuation is the interaction that couples the two free systems. The two background sections are not a solution to the interacting system, they are a reference for the method of approximation. The solution is represented by two sections that deviate (fluctuate) from the background sections. In other words, there is a fluctuation operator that represents the interaction. This situation corresponds to an interaction picture in QED. By choosing an appropiate reference frame, the background sections have the "free motion" and the fluctuation operators represent the dynamics.

We assumed that the only effect of the self interaction in the free motion of the fields is the mass parameter *m*. The remaining effects of the total physical interaction, associated with the flutuation fields, corresponds to an effective net interaction energy,

$$\tilde{e}\kappa^{\mu}e\Gamma_{\mu} - m = H \quad . \tag{13}$$

The first part of *H* is the total physical interaction. The meaning of the second part is that massive self energy is not included in the fluctuation and this method (or QED) is not adequate for calculation of mass.

It is known that the Lagrangian for the first variation of the Lagrange equations is the second variation of the Lagrangian. For both free fields, the corresponding second variation of the Lagrangian gives the free Maxwell and Dirac fields Lagrangian. For the interaction fields the second variation gives

$$\delta^2 L = \tfrac{1}{4}\operatorname{tr}(\delta\Gamma\delta J) = \tfrac{1}{4}\operatorname{tr}(\kappa^0\overline{\Psi}\gamma^{\mu}\Psi A_{\mu}) = \overline{\Psi}\gamma^{\mu}\Psi A_{\mu} \quad , \tag{14}$$

which is the standard interaction Hamiltonian in QED.

We have obtained a framework of linear operators (the prolongations of Jacobi fields) acting on states of a Banach space (sections forming a Banach space) where the operators obey the bracket commutation relations, commutators for *A* and anticommutators for $\Psi$. Furthermore, the geometric Lagrangian of the theory reduces to a Lagrangian in terms of operators which is the standard Lagrangian for QED).

From this point we can proceed using the standard language and notation of QED for any calculation we wish to carry in this approximation of the geometric theory. The result of the calculation should be interpreted physically. This is our next task, to give a statistical interpretation to fluctuations in the geometrical theory.

### *Statistical interpretation*

The physical significance of the local sections in the geometrical theory is the representation of the influence of the total universe of matter and interactions. The geometry of the theory, including the notion of corpuscles is determined by the large number of sources in the universe. Within this theory and interpretation, it is not appropriate to consider the sections to be associated with a single particle or excitation. Rather, they are associated to the global non-linear geometric effects of bulk matter and radiation.

This is not the customary situation in which physical theories are set. It is usual to postulate fundamental laws between elementary microscopic objects (particles). In order to translate the global universal geometry to the usual microscopic physics, since we associate particles to geometric excitations, we shall stablish what we call a "many excitation microscopic regime", distinguished from the situation described in the previous paragraph.

Statistics enter in our approach in a manner different than the usual one, where fundamental microscopic physical laws are postulated between idealized elementary particles and statistical analysis enters because of the difficulties that arise when combining particles to form complex systems. Here, the fundamental laws are postulated geometrically among all matter and radiation in the system ( the physical universe) and statistical analysis arises because of the difficulties and approximations in splitting the non-linear system into elementary microscopic linear fluctuation subsystems. In this holistic approach, the results associated to excitations have the statistical character of quantum theory.

We consider that we may work in two different regimes of the geometrical theory. One is the geometrical regime where there are exact non-linear equations between the local frame sections, representing matter func-tions, and the bundle connection. The other is the many excitations microscopic regime where we have linearized approximate equations between the variations of the frame and connection sections, representing particles and fields.

In the many-excitations regime, the non-linear local effects of the interaction are replaced by a linear local effects for which the systems are seen, approximately, as a collection of excitations of a background solution. The number of excitations is naturally very large and the cross interactions among them preclude an exact treatment for a single excitation. Instead it is necessary to treat the excitations as one among a large ensemble of excitations leading to a convenient and necessary use of statistical theory.

The geometric excitations form statistical ensembles of population density $n_i$. It is not possible to follow the evolution of any single one because of the previous arguments. It is absolutely necessary to use statistics in describing the evolution of the excitations. The situation is similar to that of chemical reactions, where molecules are statistically created or annihilated.

There are adequate classical statistical techniques for describing these processes. In particular we may use the theory of irreversible thermodynamics [4] to calculate the rates of reaction between different geometric excitations. The process is described by the flux density $\mathcal{J}$ characterizing the flow of excitations between two systems.

$$\mathcal{J} = \frac{dn}{dt} \quad . \tag{15}$$

It is also necessary to introduce a driving force function, $\mathcal{F}$, which is called affinity and represents differences of thermodynamic intensive parameters. In case of equilibrium between two different subsystems, both the affinity and the flux are zero.

The identification of the affinity is done by considering the rate of production of entropy s,

$$\frac{ds}{dt} = \frac{\partial s}{\partial n_k} \frac{dn_k}{dt} = \mathcal{F}^k \mathcal{J}_k \quad , \tag{16}$$

from which the affinity associated to any given excitation is

$$\mathcal{F}^k = \frac{\partial s}{\partial u} \frac{\partial u}{\partial n_k} = \frac{\mu^k}{T} \quad , \tag{17}$$

where $u$ is the energy, $T$ is the temperature and $\mu$ is the excitation potential, similar to the chemical potential.

The statistical flux is a function of the affinity and we see that the statistics of reaction of geometric excitations should depend on the classical geometric energy of the excitations. If the excitations are oscillatory, the energy depends on the excitation amplitude. Then the probability of occurrence of a single reaction event is determined by the excitation potential which in turn depends on the excitation amplitude. These is the significance of the probabilistic interpretation of quantum fields.

In some cases, for linear Markoffian systems (systems whose future is determined by their present and not their past) the flux is proportional to the affinity

$$\Delta \mathcal{J} = \left.\frac{\partial J}{\partial \mathcal{F}}\right|_0 \Delta \mathcal{F} \quad , \tag{18}$$

and the calculation of rates is simplified, indicating explicitly the dependence of the flux on the different of excitation potentials.

## Applications

In order to illustrate this statistical character of the linearized regime, we discuss the wave corpuscular duality of light and matter, typically demonstrated in Young's double slit interference experiment, within the concepts of the geometric theory. There are excitations that correspond to multiple particles and lead to Schrödinger's "entangled states". There has been a revolution in the experimental preparation of multi article entanglements [5,6,7,8,9]. In particular, consider a two-particle interferometer illustrated in figure 1. At the center there is a source of decaying particles, with a vertical extension *d*. Two collimating screens, each with a

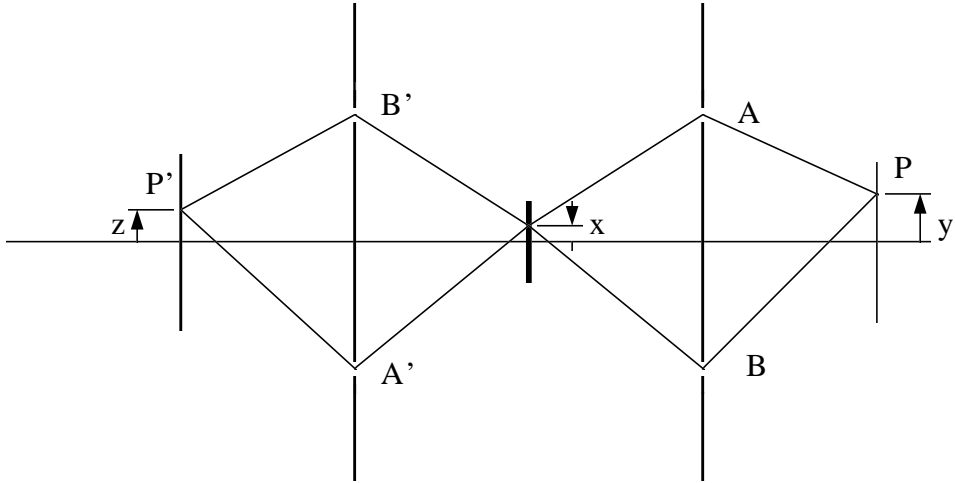

**Fig. 1**

pair of holes, offer alternative paths to a given pair of particles that are finally detected in two detecting screens

. A physical particle (f.e. a photon) is an excitation of the corresponding background field. In geometric terms, a Jacobi vector $\Psi$, associated to a variation of a section $e$ or $\omega$, represents the particle. As indicated in the previous section, it is not possible to follow the evolution of a single excitation $\Psi$. The statistical approach is to treat the radiation as a thermodynamic reservoir of excitations . This approach was used in blackbody radiation[10,11,12] and was the origin of Planck's quantum theory. In recent years, a similar idea lead to introduction of stochastic quantization [13,14], which has been shown to be equivalent to path integral quantization. Our approach is different, for example we do not introduce evolution along a fictitious time direction as done in stochastic quantization. We rely on the existence of a global nonlinear geometry which makes a practical necessity the statistical treatment of the linearized equations describing the evolution of microscopical subsytems.

The atoms in the screen are an ensemble of systems in contact with the reservoir of radiation. The equilibrium of the total system, radiation and screen, is determined by the equality of the excitation potentials associated to the geometric excitations forming the radiation. When there is no equilibrium there is a flow of excitations from the radiation to the screen. The techniques of irreversible thermodynamics relate this excitation flux density to the affinity, which depends on the differences of the associated excitation potentials. We also assume that we are dealing with a Markoffian system. This approximation has been used successfully in the quantum theory of damping [15,16] for laser systems.

In order to calculate the excitation potential we need to express the energy in terms of the number of excitations. If the electromagnetic excitation field is expressed, as usual in terms of harmonic oscillators at each point, the oscillator displacement (excitation or variation of the electromagnetic field) is a thermodynamic extensive parameter, $x$. Associated to it there is an intensive parameter $F$ that gives the energy change,

$$du = Fdx \cdot \qquad (19)$$

For linear harmonic oscillators the energy is related to the square of the displacement amplitude X

$$U = \tfrac{1}{2} K X^\dagger X \quad , \tag{20}$$

where K is a constant

As indicated previously, the Jacobi vector $\Psi$ may be decomposed in its Fourier components. The amplitudes become operators. Then the operator,

$$N = a^\dagger a \quad . \tag{21}$$

has discrete eigenvalues which determine the energy of the field. It should be clear that because of the commutation relations of the operators *a*, the energy of the harmonic oscillators corresponding to the excitations is quantized, with the same results of the quantum theory. The energy of the excitations has quanta of value *ν*, in geometric units where $\hbar$ equals 1. The radiation field has an amplitude which is modulated throughout certain space regions because of the interference pattern produced by the waves. This means that the average energy is not homogeneously distributed but concentrated in certain regions. We should consider a local excitation potential defined within domains of volume, determined by a correlation distance where the density of energy *u* may taken as constant. Then the excitation potential is

$$\mu = \frac{\partial u}{\partial n} \quad , \tag{22}$$

where *n* is the eigenvalue of *N*.

In the standard Young´s experiment the waves arriving at a point in the screen have the well known phase difference $\phi$ that allow us to write for the resultant amplitude,

$$\Psi = \Psi_0 e^{i\phi} + \Psi_0 e^{-i\phi} \quad . \tag{23}$$

When the amplitude $\Psi_0$ is expressed in terms of the Fourier components *a*, the energy of the excitation leads to the expression for the potential,

$$\mu = 4\varepsilon \cos\phi = 4\varepsilon \cos\left(\frac{\pi l}{\lambda}\sin\alpha\right) \quad . \tag{24}$$

In our case we have to consider an *m*-particle wave. To clarify the situation and avoid confusion with the concept of particle, we define an *m*-corpuscle excitation as the tensor product representation of *m* fundamental (1-corpuscle) representations. The Jacobi excitation field amplitude is then an *m*-product representation.

In this language, we consider the case of a 2-corpuscle excitation experiment, in particular the case of two Young's experiments side by side. We not only have a correlation betweeen alternative paths through A and B for the right corpuscle but also correlations among paths A or B and A' or B' for the 2-corpuscle excitation formed by the product of the two 1-corpuscle excitations. If a particle decays at height *x* above the center line, a particle may be detected in the right screen at height *y* and another in the left screen at height *z*. The phase difference $\varphi$ for the right side particle has a contribution from the screen angle

$$\phi_P = \frac{\pi l y}{\lambda r} \quad , \tag{25}$$

and another from the source angle,

$$\phi_P = \frac{\pi l x}{\lambda r} \quad , \tag{26}$$

leading to, for small y an z,

$$\phi_P = \frac{\pi\theta}{\lambda}(x+y) \quad , \tag{27}$$

where $\theta$ is the angle subtended by the two holes. A similar expression is obtained for the left side particle,

$$\phi_{P'} = \frac{\pi\theta}{\lambda}(x+z) \quad . \tag{28}$$

The amplitude of the Jacobi excitation, which is a product representation of the two excitations, is

$$\Psi \propto \frac{a}{d}\int_{-d/2}^{+d/2} dx \cos\left[\frac{\pi\theta}{\lambda}(x+y)\right]\cos\left[\frac{\pi\theta}{\lambda}(x+z)\right] \quad (29)$$

If $d$ is much smaller than $\lambda/\theta$ the integral gives the product of two Young's patterns as it should. If, on the contrary, $d$ is much larger than $\lambda/\theta$, this integral gives,

$$\Psi \propto \frac{a}{2}\cos\left(\frac{\pi\theta}{\lambda}(z-y)\right) \ . \quad (30)$$

We obtain for the excitation potential

$$\mu \propto \left[\cos\left(\frac{\pi\theta}{\lambda}(z-y)\right)\right]^2 \ . \quad (31)$$

The probability for the simultaneous detection of particles at P an P' is given by this excitation potential. There are regions of high and low probabilities because of the interference of the excitation amplitudes. These cases just illustrated for 1-particle and 2-particle interferometers give a general characteristic of section excitation in the geometrical theory. There are no exact equations for physical situations with a single excitation. Any real physical situation in the microscopic linear regime of the theory requires the use of statistics. The outcome of microscopic transition experiments depends on the excitation potentials of the systems. The requirement of statistics necessarily leads to the probability interpretation of quantum theory. If, within a particular experimental setup, we can physically distinguish between two excitation states, there no room to apply statistics and there follows the absence of the interference pattern. This is the content of Feynman's statement about experimentally indistinguishable alternatives [17]. Fundamentally the quantum statistics are the classical statistics of section excitations in this physical geometry. The objections raised by Einstein to the probabilistic interpretation are resolved automatically [18,19].

## Conclusion

Excitations of a physical geometry were used to represent the theory of quantum electrodynamics. The connection excitations and the frame excitations reduce, respectively, to the electromagnetic field operator and electron field operator. Because of the inherent geometric algebraic structure this operators obey the standard commutation rules of QED. Use of harmonic analysis introduces creation and annihilation operators for the associated excitation waves. The energy of the connection excitations is $\hbar\nu$.

The nonlinear geometric equations apply to the total universe of matter and radiation. If we work with excitations, this implies the need to use statistical theory when considering the evolution of microscopic subsystems. The use of classical statistics, in particular techniques of irreversible thermodynamics, determine that the probability of absorption or emission of a geometric excitation is a function of the classical energy density, which determines the excitation (chemical) potential.

Emission and absorption of geometric excitations imply discrete changes of certain physical variables, but with a probability determined by its wave energy density. Hence, this geometric theory does not contradict the fundamental aspects of quantum theory. On the contrary, it offers a geometrical representation for the existence of discrete quanta of energy, spin, electric charge and magnetic flux.

## References

1 G. González-Martín, Gen. Rel. Grav. 23, 827 (1991); G. González-Martín, Physical Geometry, (Universidad Simón Bolívar, Caracas) (2000), Electronic copy posted at http:\\prof.usb.ve\ggonzalm\invstg\book.htm
2 G. González-Martín, Gen. Rel. Grav. 24, 501 (1992).
3 G. González-Martín, Phys. Rev. D 35, 1225 (1987).
4 H. B. Callen, Thermodynamics, (J. Wiley & Sons, New York), p. 289 (1960).
5 M. A. Horne, A. Zeilinger, in Proc. Symp. on Foundations of Modern Physics, P. Lahti, P. Mittelstaedt, eds.,

(World Science, Singapore), p. 435 (1985).
6C. O. Alley, Y. H. Shih, Proc. 2nd Int. Symp. on Foudations of Quantum Mechanics in the Light of New Technology, M. Namiki et al, eds. (Phys. Soc. Japan, Tokyio), p. 47 (1986).
7M. A. Horne, A. Shimony, A Zeilinger, Phys. Rev. Lett. 62, 2209 (1989).
8R. Ghosh, L. Mandel, Phys. Rev. Lett. 59, 1903 (1987).
9Y. H. Shih, C. O. Alley, Phys Rev. Lett. 61, 2921 (1988).
10M. Planck, Verh. Dtsch. Phys. Gesellschaft, 2, 202 (1900).
11S. Bose, Z. Physik 26, 178 (1924).
12A. Einstein, Preuss. Ak. der Wissenschaft, Phys. Math. Klasse, Sitzungsberichte, p, 18 (1925).
13G. Parisi, Y. S Wu, Sc. Sinica, 24, 483 (1981).
14P. H. Damgaard, H. Huffel, Physics Reports 152, 229 (1987).
15W. H. Louisell, L. R. Walker, Phys. Rev. 137B, 204 (1965).
16W: H. Louisell, J. H. Marburger, J. Quantum Electron. QE-3, 348 (1967).
17R. P. Feynman, The Feyman Lectures on Physics, Quantum Mechanics. (Addison Wesley, Reading), p.3-7 (1965).
18A. Einstein, B. Podolsky, N. Rosen, Phys. Rev. 47, 777 (1935).
19N. Bohr, Phys. Rev. 48, 696 (1936).